\documentclass[11pt]{article}
\usepackage[margin=1in]{geometry}
\usepackage{booktabs}
\usepackage{amsmath}
\usepackage[hidelinks]{hyperref}
\usepackage{xcolor}
\usepackage{listings}
\usepackage{tikz}
\usetikzlibrary{arrows.meta,positioning,fit,calc}
\title{TGMS: An Agent-Native Bi-Temporal Graph Management System\\
\large Validated Temporal Operators and Trace-Grounded Answer Checking}
\author{Xiaofei Zhang\\University of Memphis\\\texttt{xiaofei.zhang@memphis.edu}}
\date{July 2026 \;\textperiodcentered\; System Description Preprint (v3)
\;\textperiodcentered\; \url{https://github.com/zxf-work/tgms}\thanks{This
preprint accompanies TGMS v0.3.0: development results, the frozen-test
campaign (thresholds recorded in the repository's dated decision log
before measurement), and post-campaign scale, fair-baseline, and
portability studies.}}

\begin{document}
\maketitle

\begin{abstract}
Temporal graph question answering requires exact composition over time
and, when records are corrected, reconstruction of prior belief states.
Large language model (LLM) agents are unreliable at identifier
grounding, arithmetic, and interpreting partial query results. We
present TGMS, an agent-native bi-temporal graph system that assigns the
model two bounded roles: constructing a plan over a fixed operator
interface and verbalizing the result. A static verifier checks plan
structure, identifier grounding, declared output fields, temporal
arguments, and estimated cost. Thirteen deterministic temporal
operators execute over valid-time and transaction-time data and produce
content-addressed traces with completeness metadata. A claim verifier
checks the final answer against those traces.

On a frozen 94-task CollegeMsg test split run with three seeds, TGMS
using Qwen2.5-14B-Instruct-AWQ obtains 0.408 normalized typed-answer
accuracy, compared with 0.106 for vector-RAG, 0.064 for static-graph
RAG, and 0.152 for text-to-Cypher. On correction probes, TGMS obtains
0.897; the two latest-state baselines obtain zero, while vector-RAG
obtains 0.154 by answering current-belief cases. The main result
transfers to a second communication corpus and a synthetic
temporal-pattern suite, although complex multi-operator pattern plans
remain beyond the 14B planner. Before gating, 21 of 270 reporter
answers (7.8\%) contain at least one unsupported gated claim;
verification reduces this to 0 of 268 at a 1.0-percentage-point
reduction in overall answer accuracy. Mutation experiments show that
output contracts and evidence-completeness tracking catch errors that
value-only checking misses. In a separate full-precision serving
configuration, accuracy rises markedly with model scale
(0.138/0.340/0.628 at 7B/14B/32B) while the tested baselines change
little, and increasing vector-RAG's retrieval breadth to a long-context
configuration does not improve it. These results suggest that
agent-facing database interfaces should make belief time, result
contracts, and evidence completeness explicit.
\end{abstract}

\section{Introduction}
Temporal graphs arise in communication logs, transaction records, and
evolving knowledge bases. They create two difficulties for LLM agents
that are less visible in static text collections.

The first difficulty is temporal composition. Consider the question:
``Among accounts reachable from $X$ in February, how many cyclic
triangles closed within one day?'' Answering it requires
time-respecting reachability~\cite{wu2014path}, followed by
$\delta$-temporal motif counting~\cite{paranjape2017motifs}. A retrieval
pipeline that serializes edges into text may omit the exact structure
needed by the second step.

The second difficulty is belief revision. A graph may change because the
world changed, or because an earlier record was wrong. These cases are
different. Answering ``what did we believe on March~1?'' requires both
valid time and transaction time~\cite{snodgrass1999}. A latest-state
snapshot cannot reconstruct this distinction after a correction has been
applied.

LLMs introduce a separate set of risks. They may perform arithmetic
incorrectly, invent identifiers, or report values that do not appear in
the evidence.

These problems share one cause: conventional database interfaces assume
a competent client, whereas an LLM is an approximate planner and
reporter. TGMS therefore places the LLM outside the trusted computing
boundary. The model chooses and composes operations and verbalizes
their results, while the database owns temporal semantics, identifier
grounding, arithmetic, bounded execution, provenance, and claim
checking. We call a database interface \emph{agent-native} when it is
designed for a fallible machine planner: operations have
machine-readable input and output contracts, explicit bounds and costs,
deterministic result identities, and evidence metadata that supports
automated checking.

This paper makes four contributions, each subordinate to that thesis:
\begin{itemize}
    \item \textbf{Belief-state-preserving storage:} a bi-temporal
    property graph substrate with explicit assertion, retraction, and
    correction, backed by an append-only event log that replays to
    byte-identical state across storage backends.
    \item \textbf{A machine-checkable computation interface:} a fixed
    algebra of thirteen temporal operators with typed input \emph{and
    output} contracts, deterministic results, pagination, and
    pre-execution cost guards.
    \item \textbf{Evidence-aware execution and answer verification:} a
    static plan verifier, a deterministic executor that produces
    content-addressed traces with completeness metadata, and a claim
    verifier that gates final answers against those traces.
    \item \textbf{Empirical evidence about when and why this separation
    helps:} frozen test suites over two real communication networks and
    one synthetic temporal-pattern workload, measuring answer accuracy,
    correction handling, plan validity, unsupported claims, verifier
    fault detection, model-scale behavior, baseline sensitivity to
    context limits, and operator latency.
\end{itemize}

\paragraph{Evaluation status and chronology.}
We first used a 22-task development split to finalize the operator
interface, prompts, repair protocol, and acceptance criteria; the
acceptance thresholds were recorded in the repository's dated decision
log and committed evaluation driver before measurement (an internal
dated record, not an external registry). We then froze the evaluation
protocol and ran the reported test campaign on 290 unique tasks, with
three seeds for CollegeMsg. Post-campaign studies examine model scale,
quantization, longer-context retrieval, and cross-family portability;
these are reported separately from the frozen primary evaluation
(Section~\ref{sec:sensitivity}).

\subsection{System overview and trust boundary}
\begin{figure}[t]
\centering
\resizebox{\textwidth}{!}{
\definecolor{figamber}{RGB}{214,138,0}
\definecolor{figamberfill}{RGB}{255,247,224}
\definecolor{figgreen}{RGB}{46,125,50}
\definecolor{figgreenfill}{RGB}{232,245,233}
\definecolor{figblue}{RGB}{21,101,192}
\definecolor{figbluefill}{RGB}{227,242,253}
\definecolor{figgray}{RGB}{110,116,124}
\definecolor{figgrayfill}{RGB}{245,246,247}
\definecolor{figbound}{RGB}{26,86,196}
\begin{tikzpicture}[
  every node/.style={font=\small},
  box/.style={rounded corners=7pt, align=center, inner ysep=8pt,
              inner xsep=8pt, line width=1.1pt},
  llm/.style={box, draw=figamber, fill=figamberfill},
  vrf/.style={box, draw=figgreen, fill=figgreenfill},
  exe/.style={box, draw=figblue, fill=figbluefill},
  gry/.style={box, draw=figgray, fill=figgrayfill},
  badge/.style={font=\scriptsize\bfseries, text=white, rounded corners=3pt,
                inner xsep=5pt, inner ysep=2.5pt},
  ttl/.style={font=\small\bfseries},
  lbl/.style={font=\scriptsize, text=black!60},
  arr/.style={-{Stealth[length=2.8mm]}, line width=1.1pt, black!75},
  darr/.style={arr, dashed, black!55},
  leg/.style={rounded corners=3pt, line width=1pt, minimum width=7mm,
              minimum height=3.6mm}]

\node[llm, text width=33mm] (plan) at (0,0)
  {\textbf{Planner}\\[2pt] {\scriptsize emits the plan IR\\ (JSON DAG + \$refs)}};
\node[badge, fill=figamber] at ($(plan.north west)+(0.55,0)$) {LLM};

\node[llm, text width=33mm] (rep) at (0,-6.6)
  {\textbf{Reporter}\\[2pt] {\scriptsize writes the answer object\\ (claims cite evidence)}};
\node[badge, fill=figamber] at ($(rep.north west)+(0.55,0)$) {LLM};

\node[vrf, text width=44mm] (ver) at (7.1,0)
  {\textbf{Static verifier}\\[2pt] {\scriptsize grounding $\cdot$ output contracts\\
    temporal sanity $\cdot$ cost}};
\node[badge, fill=figgreen] at ($(ver.north west)+(0.7,0)$) {CHECK};

\node[exe, text width=38mm] (exec) at (12.3,0)
  {\textbf{Executor}\\[2pt] {\scriptsize deterministic\\
    content-addressed $\cdot$ truncation taint}};

\node[gry, text width=88mm, align=center] (stack) at (9.65,-3.1)
  {\textbf{Operator layer}~{\scriptsize (in-process router / MCP)}\\[1pt]
   {\scriptsize O1--O13: typed $\cdot$ deterministic $\cdot$ bounded $\cdot$
    bi-temporal (\texttt{as\_of\_tt}) $\cdot$ cost-guarded}\\[4.5pt]
   \textcolor{figgray}{\rule{80mm}{0.4pt}}\\[3.5pt]
   \textbf{Bi-temporal substrate}\\[1pt]
   {\scriptsize versions over valid $\times$ transaction time $\cdot$
    write-ahead event log $\cdot$ temporal CSR}};

\node[vrf, text width=76mm] (cver) at (9.05,-6.6)
  {\textbf{Claim verifier}\\[2pt] {\scriptsize counts $\cdot$ entities $\cdot$ orderings
    $\cdot$ patterns re-checked against trace digests\\
    unsupported claims are gated out of the answer}};
\node[badge, fill=figgreen] at ($(cver.north west)+(0.7,0)$) {CHECK};

\node[draw=figbound, dashed, line width=1.3pt, rounded corners=10pt,
      inner sep=11pt, fit=(ver)(exec)(stack)(cver)] (tcb) {};
\node[badge, fill=figbound, font=\scriptsize\bfseries]
  at ($(tcb.north west)+(1.05,0)$) {TRUSTED};

\node[align=center, text width=26mm] (out) at (16.9,-6.6)
  {\textbf{Verified answer}\\[1pt]{\scriptsize every claim cites trace digests}};
\node[draw=black!70, line width=1pt, rounded corners=2pt, minimum width=9mm,
      minimum height=11mm] (doc) at ($(out.north)+(0,0.85)$) {};
\draw[figgreen, line width=1.4pt, line cap=round]
  ($(doc.center)+(-0.14,0.0)$) -- ($(doc.center)+(-0.03,-0.12)$)
  -- ($(doc.center)+(0.16,0.14)$);

\draw[arr] (plan) -- node[lbl, below, align=center, pos=0.38]
  {plan\\[-1pt] {\tiny\ttfamily \{"steps":[...]\}}} (ver);
\draw[darr] (ver.north) to[bend right=14]
  node[lbl, above=3pt, pos=0.82, xshift=-7mm]{repair payloads (E\_SCHEMA / E\_COST / ...)}
  (plan.north);
\draw[arr] (ver) -- node[lbl, above]{valid} (exec);
\draw[arr] (exec.south) -- node[lbl, right]{operator calls}
  (exec.south |- stack.north);
\draw[arr, rounded corners=8pt]
  (exec.east) -- ++(0.7,0) |- node[lbl, right, pos=0.25]{trace} (cver.east);
\draw[arr] (rep) -- node[lbl, below]{answer object} (cver);
\draw[darr, rounded corners=8pt]
  (stack.west) -| node[lbl, left, pos=0.75]{trace summaries} (rep.north);
\draw[arr, rounded corners=8pt] (cver.south) -- ++(0,-0.55) -|
  node[lbl, below, pos=0.22]{gated claims} (out.south);

\begin{scope}[shift={(1.2,-9.4)}]
  \node[leg, draw=figamber, fill=figamberfill] (l1) at (0,0) {};
  \node[right=1.5mm of l1, font=\scriptsize] {LLM (untrusted)};
  \node[leg, draw=figgreen, fill=figgreenfill] (l2) at (3.6,0) {};
  \node[right=1.5mm of l2, font=\scriptsize] {verification};
  \node[leg, draw=figblue, fill=figbluefill] (l3) at (6.8,0) {};
  \node[right=1.5mm of l3, font=\scriptsize] {execution};
  \node[leg, draw=figgray, fill=figgrayfill] (l4) at (9.8,0) {};
  \node[right=1.5mm of l4, font=\scriptsize] {data layer};
  \node[leg, draw=figbound, dashed] (l5) at (12.8,0) {};
  \node[right=1.5mm of l5, font=\scriptsize] {trust boundary};
\end{scope}
\end{tikzpicture}}
\caption{TGMS as a trust boundary. The planner and reporter (the LLM)
are outside the trusted region; the static verifier, operator engine,
bi-temporal store, execution trace, and claim verifier are inside it.
Model outputs cross the boundary only through checks.}
\label{fig:arch}
\end{figure}
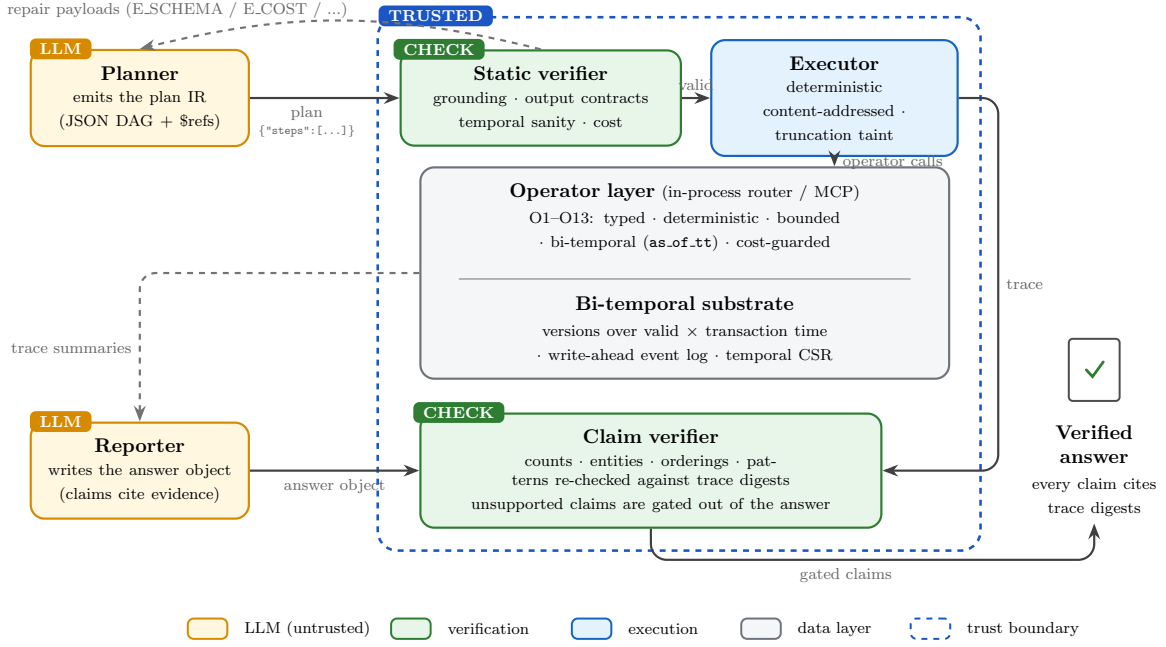

Figure~\ref{fig:arch} shows the architecture as a trust boundary. The
LLM appears exactly twice --- as planner and as reporter --- and both
roles sit outside the boundary: nothing the model produces is executed
or emitted unchecked. Inside the boundary, deterministic components own
everything that must be true: the static verifier admits or rejects
plans; the operator engine computes over the bi-temporal store; the
executor records content-addressed evidence; and the claim verifier
checks the final answer against that evidence. The remainder of the
paper follows this decomposition: the substrate
(Section~\ref{sec:substrate}), the operator interface
(Section~\ref{sec:operators}), and planning, execution, and
verification (Section~\ref{sec:pev}).

\section{Bi-temporal substrate}
\label{sec:substrate}
Each logical node and edge has a stable identity and one or more
versions. A version carries a half-open valid-time interval
$[vt_s, vt_e)$ and a half-open transaction-time interval $[tt_s, tt_e)$.
TGMS stores time as int64 epoch microseconds and uses $2^{62}$ as the
open-ended upper bound. A snapshot $G(t,tt)$ contains the versions whose
valid-time intervals contain $t$ and whose transaction-time intervals
contain $tt$.

The write API provides three bi-temporal operations. \texttt{assert}
records a new belief and carves its valid-time interval out of any
overlapping prior belief of the \emph{same logical identity};
independently coexisting facts use distinct identities, so within one
identity the believed valid-time intervals are always pairwise
disjoint. \texttt{retract} closes a fact's valid time (the world
changed; the earlier belief remains correct for its interval).
\texttt{correct} closes the \emph{transaction time} of an erroneous
version and records its replacement (we were wrong; the record of the
error is preserved). TGMS also provides a bulk-ingest path for
instantaneous event streams.

A minimal worked example. At transaction time $tt_1$ we assert that
edge $e$ has weight $p{=}1$ over valid time $[0, 100)$: one version,
$vt{=}[0,100)$, $tt{=}[tt_1, \infty)$. At $tt_2$ we \emph{correct} the
weight to $p{=}2$ for $vt{=}[40, 60)$: the original version's
transaction interval closes at $tt_2$, and three new versions open ---
$p{=}1$ over $[0,40)$, $p{=}2$ over $[40,60)$, and $p{=}1$ over
$[60,100)$, all with $tt{=}[tt_2,\infty)$. The current snapshot at
$vt{=}50$ returns $p{=}2$; the same query pinned to
$\texttt{as\_of\_tt}{=}tt_1$ returns $p{=}1$ --- the belief before the
correction --- and continues to do so forever.

All writes pass through an append-only write-ahead event log. The log
records every \emph{attempted} batch together with its deterministic
outcome: only successfully committed batches contribute to the
materialized graph state, and replay reproduces both the successful
state transitions and the recorded failures, skipping the latter
exactly as the original writer did. In this sense the log is the source
of truth for state and audit trail for attempts. Replaying it into a
backend reproduces the same store digest. We test this property across K\`uzu~\cite{kuzu2023} and
DuckDB~\cite{duckdb2019} adapters. Update semantics are implemented once,
above a small adapter interface, so the backends share the same behavior.
A hybrid logical clock provides strictly increasing transaction times.
Failed batches remain in the log. Replay reproduces the failure and skips
the batch in the same way. Version rows also reserve
\texttt{source} and \texttt{provenance\_ref} fields so that future agent
write-back will not require a schema migration.

Property-based tests generate random sequences of assertions,
retractions, and corrections. They check two main invariants. First, the
versions believed for one identity have pairwise-disjoint valid-time
intervals at every historical transaction time. Second, results pinned to
a past \texttt{as\_of\_tt} remain byte-identical after later corrections.
We call the second property \emph{bi-temporal immutability}. Enforcing it
required two details. Result digests must exclude metadata that describes
the current belief state. Returned versions must also hide
belief-closure timestamps that occur after the requested transaction
time.

\section{A fixed operator algebra}
\label{sec:operators}
TGMS exposes the thirteen operators in Table~\ref{tab:ops}. The
interface is \emph{fixed}: plans may invoke only operators from the
registered set; we do not claim algebraic completeness. The set covers
four workload families --- historical state (O1--O3), temporal
traversal (O4--O5), temporal patterns (O6--O9), and evolution analysis
(O10--O11) --- plus entity grounding (O12) and deterministic
post-processing (O13). It is not intended to be complete; the
evaluation measures which task families are expressible and which still
exceed the planner or the interface. The \texttt{compute} operator
keeps arithmetic out of the LLM, and \texttt{resolve\_entities} is the
only operator that may introduce an identifier not already present in
the task.

\begin{table}[t]
\centering\small
\caption{TGMS operator algebra. Each operator also accepts
\texttt{as\_of\_tt}, \texttt{limit}, and \texttt{cursor}, and declares its
output fields.}
\label{tab:ops}
\begin{tabular}{llp{8.2cm}}
\toprule
 & Operator & Semantics \\
\midrule
O1 & \texttt{entity\_history} & Returns the version history of a node under a selected belief state. \\
O2 & \texttt{snapshot\_subgraph} & Returns a $k$-hop neighborhood, with $k \le 3$, from snapshot $G(t,tt)$. \\
O3 & \texttt{diff\_snapshots} & Returns additions, removals, and changes between $G(t_1)$ and $G(t_2)$. \\
O4 & \texttt{temporal\_reachability} & Computes earliest arrival over time-respecting paths, with exact handling of wait limits. \\
O5 & \texttt{temporal\_paths} & Returns up to $k \le 20$ node-simple, time-respecting paths. \\
O6/O7 & \texttt{count/find\_temporal\_motifs} & Counts or returns exact $\delta$-temporal motifs~\cite{paranjape2017motifs}, ordered by $(t,\textit{eid})$. \\
O8 & \texttt{graph\_metric\_timeseries} & Returns bucketed node, edge, degree, and reciprocity statistics. \\
O9 & \texttt{burst\_detection} & Detects unusual buckets with a rolling z-score or trailing-median rule. \\
O10 & \texttt{neighborhood\_evolution} & Returns neighbors gained or lost and a degree time series. \\
O11 & \texttt{co\_active} & Applies an Allen-relation interval join to two edge selections. \\
O12 & \texttt{resolve\_entities} & Resolves names or user identifiers to graph entities. \\
O13 & \texttt{compute} & Applies count, sum, min, max, top-$k$, filter, or interval-relation operations to prior outputs. \\
\bottomrule
\end{tabular}
\end{table}

Five rules apply to every operator.

\paragraph{Typed contracts.}
Inputs and outputs are validated against JSON Schemas generated from one
registry. The same registry generates the tool definitions shown to the
agent.

\paragraph{Deterministic results.}
The same store state and arguments produce the same canonical output and
the same SHA-256 \texttt{result\_digest}. Floating-point values are
canonicalized before both serialization and threshold comparisons. This
keeps the engine and its test oracle consistent.

\paragraph{Bounded execution.}
Two distinct bounds apply. Result \emph{cardinality} is bounded through
\texttt{limit}/\texttt{cursor} pagination. Computational \emph{work}
is bounded separately through pre-execution cost estimation and refusal
--- a motif count can return one number after an expensive search, so
pagination alone would not bound it. The rejection includes concrete
ways to narrow the request, which
the planner may use during repair.

\paragraph{Bi-temporal semantics.}
Every operator accepts \texttt{as\_of\_tt}. A plan can therefore query a
past belief state without changing the operator interface.

\paragraph{Declared outputs.}
Each operator lists its output fields in the registry. The static verifier
rejects references to fields that do not exist. Section~\ref{sec:findings}
shows why this check matters for small models.

Correctness testing is a release gate. Each operator is compared with an
independent brute-force implementation on 500 randomized combinations of
stores and arguments. We also test metamorphic properties, including diff
composition and bi-temporal immutability. These tests exposed problems in
both the code and the specification. One test found a collision in the
original version identifier. Another showed that a greedy reachability
rule under a maximum-wait constraint depended on processing order. TGMS
now uses an exact multi-label search over
$(\text{node},\text{arrival})$ states.

\section{Planner--Executor--Verifier}
\label{sec:pev}
A TGMS plan is a small JSON DAG. Each step calls one operator. References
between steps use a limited \texttt{\$ref} language with dotted fields,
\texttt{rows[i]}, and \texttt{rows[*].field} projections. The language
does not allow general expressions. An \texttt{answer\_spec} identifies
the step and field that contain the final result.

Before execution, a static verifier checks the plan schema, acyclicity,
reference scope, temporal arguments, output fields, and estimated cost.
It also applies a grounding rule. A literal identifier is allowed only if
it appears in the task input. Other identifiers must be produced by a
prior step and passed through \texttt{\$ref}. This prevents plans from
introducing identifiers that were neither supplied nor resolved.
Rejections return structured violations. The planner may revise a plan up
to three times. We report both first-attempt validity and success after
repair.

The executor runs the DAG deterministically and records
content-addressed results. Re-executing the same plan over the same store
state reproduces the same result digests. The executor also propagates
\emph{truncation taint}. If a step reads a truncated result page, that
step and all dependent steps are marked as using incomplete evidence.

A reporter LLM produces a typed answer object. Each claim cites one or
more evidence steps. The claim verifier checks counts and values against
named trace fields. It checks entity mentions against the cited content,
re-evaluates Allen relations for ordering claims, and re-executes
operators for temporal pattern claims. A claim that cites truncated or
tainted evidence can be rated no higher than \emph{weakly supported}.
Count, value, entity, and ordering claims are currently gated before the
answer is emitted. Pattern checks are reported but are not yet used to
block output.

\subsection{Extensions outside the evaluated pipeline}
Two components ship with TGMS but are not exercised by the primary
evaluation; we summarize them for completeness. First, a
\emph{threat-model} defense: stored graph content is treated as data
rather than instructions --- strings inserted into prompts are escaped,
length-limited, and placed inside explicit data fences, and red-team
fixtures for this boundary run in continuous integration. Second, an
\emph{evolution memory}: operator-computed facts over weekly windows
are summarized by an LLM, stored only when every numeric statement
matches the computed values, and stamped with their creation
transaction time. A later correction quarantines any note whose
valid-time window overlaps the corrected interval. Quarantined notes
are excluded from planner and reporter context and cannot be cited;
final claims must still resolve to operator traces --- the memory
supplies hints, never evidence.

\section{Evaluation}
\label{sec:eval}

\subsection{Questions and experimental design}
The evaluation tests four hypotheses:
\textbf{H1} transaction-time storage enables correction queries that are
unavailable from latest-state inputs;
\textbf{H2} operator output contracts improve plan validity and
execution success;
\textbf{H3} completeness-aware verification reduces unsupported output;
\textbf{H4} the interface remains usable across data scales and model
configurations.
We first used a 22-task development split to finalize the interface,
prompts, repair protocol, and acceptance thresholds; the protocol was
then frozen (task generator, splits, and split SHAs recorded in the
repository's dated decision log) before any test-split measurement, and
task generation was not altered afterwards. All systems use
temperature~0, the same model checkpoint, the same seeds, the same
repair budget, and the same typed answer format.

\subsection{Data, frozen suites, systems, and metrics}
\label{sec:setup}

\paragraph{Task construction.}
Tasks are program-generated with engine-computed gold: oracle plans are
executed by TGMS itself, so no LLM-generated label is ground truth. The
generator defines 17 semantic templates (10 single-operator temporal
questions, T1; 3 evolution questions, T3; 4 multi-step analytical
questions, T4), each with three hand-written paraphrases; every
instantiated task samples template parameters (entities, windows,
thresholds) from the store and uses one paraphrase. Instances whose
oracle execution fails or degenerates are discarded. Correction probes
are constructed by injecting a correction and generating a
\emph{pair} of separately scored questions over the same logical fact
--- one asking for the belief before the correction, one for the
current belief --- keeping only pairs whose two gold answers differ.
The synthetic suite adds planted-pattern mining tasks (T2) whose gold
comes from the generator's manifest. Tasks are split 20/80
(development/test), stratified by family; because the split cuts
across probe pairs, a split may contain one or both halves of a pair.

\begin{table}[t]
\centering\small
\caption{Frozen suites. CollegeMsg test answer types: 42 count,
34 value, 14 interval, 4 entity-set. The CollegeMsg test split
contains 13 probe \emph{questions}: 7 before-correction halves and 6
current-belief halves.}
\label{tab:data}
\begin{tabular}{lcccccc}
\toprule
suite & dev & test & T1 & T2 & T3 & T4 / probes \\
\midrule
CollegeMsg & 22 & 94 & 48 & --- & 20 & 13 / 13 \\
email-EU & 22 & 94 & 48 & --- & 20 & 13 / 13 \\
synthetic & 24 & 102 & 48 & 8 & 20 & 13 / 13 \\
\bottomrule
\end{tabular}
\end{table}

Because the operator set and the workload were co-designed, the study
does not establish coverage of independently collected temporal-graph
questions; the frozen protocol limits tuning to the test sets but not
this closed-world alignment (Section~\ref{sec:limits}).

\paragraph{Systems and the information each receives.}
Table~\ref{tab:baselineinfo} makes the input representations explicit,
because the correction probes are a \emph{capability} comparison ---
they demonstrate the value of retaining transaction-time history ---
while the non-probe tasks compare the reliability of different planning
and data-access interfaces over comparable information.

\begin{table}[t]
\centering\small
\caption{What each system can see. ``valid time'' for vector-RAG means
timestamps serialized into chunk text; no baseline retains
transaction-time history.}
\label{tab:baselineinfo}
\begin{tabular}{lp{3.6cm}cccc}
\toprule
system & input representation & valid t. & trans.\ hist. & exact ops & bounded \\
\midrule
TGMS & bi-temporal versions & yes & yes & yes & yes \\
Vector-RAG & retrieved chunks of serialized event sentences & in text & no & no & yes \\
Static-graph RAG & 2-hop edge lists, latest snapshot & current only & no & no & yes \\
Text-to-Cypher & latest-state property graph, timestamp property & as property & no & query engine & yes \\
\bottomrule
\end{tabular}
\end{table}

We also run a no-verifier TGMS ablation whose raw reporter answers are
scored by the claim verifier acting as a measurement instrument without
gating.

\paragraph{Metric: normalized typed-answer accuracy.}
Scoring compares the typed final answer, not prose. Count and value
answers must agree within $10^{-9}$ after float canonicalization;
entity-set answers require set equality of identifiers; interval
answers count as correct when interval intersection-over-union is at
least 0.5; remaining kinds use canonical-JSON equality. Because of the
interval rule we call the pooled measure \emph{normalized typed-answer
accuracy} rather than exact match; counts and values are exact.
One known laxity: entity identifiers are collected from
\texttt{uid}/\texttt{src}/\texttt{dst} fields at any nesting depth,
which can accept an answer with the right identifiers in the wrong
structure. Gold answers are computed by the engine under the same
normalization.

\paragraph{Unsupported-claim rate.}
An answer is \emph{unsupported} if at least one emitted gated claim
(count, value, entity, or ordering; pattern claims are checked but not
yet gated) is rejected by the claim verifier. We report the
answer-level rate with explicit denominators, together with
\emph{coverage} (fraction of tasks for which an answer with at least
one gated claim is emitted) and \emph{conditional accuracy} (accuracy
among emitted answers), because a zero unsupported rate can arise
either from correct supported answers or from abstention.

\subsection{Primary frozen-test results}
\label{sec:frozen}
Table~\ref{tab:frozen} reports the campaign: frozen splits, three seeds
for CollegeMsg, Qwen2.5-14B-Instruct-AWQ served by vLLM~0.11 on a
single 24\,GB Turing GPU.

\begin{table}[t]
\centering\small
\caption{Frozen-test campaign. Model: Qwen2.5-14B-Instruct-AWQ (AWQ
4-bit), vLLM 0.11, 24\,GB Turing GPU; CollegeMsg pools three seeds
(per-seed spread one task), other suites one seed. Paired bootstrap
over CollegeMsg tasks (10k resamples): TGMS $-$ vector-RAG $+0.30$
$[+0.21,+0.40]$; $-$ static-graph RAG $+0.34$ $[+0.25,+0.44]$; $-$
text-to-Cypher $+0.26$ $[+0.15,+0.36]$.}
\label{tab:frozen}
\begin{tabular}{lcccc}
\toprule
 & TGMS & Vector-RAG & Static-graph & Text-to-Cypher \\
\midrule
CollegeMsg (94$\times$3) & \textbf{0.408} & 0.106 & 0.064 & 0.152 \\
email-EU (94) & \textbf{0.309} & --- & 0.053 & 0.106 \\
synthetic (102) & \textbf{0.314} & --- & 0.029 & 0.157 \\
probes, CollegeMsg (13$\times$3) & \textbf{0.897} & 0.154 & 0.000 & 0.000 \\
probes, email-EU (13) & \textbf{0.846} & --- & 0.000 & 0.000 \\
\bottomrule
\end{tabular}
\end{table}

\paragraph{Correction probes (H1).}
TGMS answers 0.897 of probe questions on CollegeMsg and 0.846 on
email-EU. The two latest-state baselines score exactly zero;
vector-RAG's 0.154 comes entirely from current-belief halves, whose
answers a latest-state input can contain. The before-correction halves
require the earlier belief state, which no baseline input preserves ---
a representational distinction, not a planning-quality one
(Table~\ref{tab:baselineinfo}).

\paragraph{Answer support, coverage, and the cost of gating (H3).}
Of 270 raw reporter answers measured on CollegeMsg (the no-verifier
ablation; 282 tasks, 12 runs produced no measurable report), 21
contained at least one unsupported gated claim: an answer-level rate of
7.8\%. With gating, 0 of 268 measured emitted answers contained an
unsupported gated claim. Gating cost 1.0 percentage point of overall
accuracy (0.418 ungated vs.\ 0.408 gated). The gated system's coverage
was 199/282 (0.706) with conditional accuracy 0.548 among emitted
answers --- so part of the zero unsupported rate is bought by
abstention, and we report both sides of that trade. These guarantees
extend only to gated claim types, not to pattern claims, approximate
operators, write-back, or adversarial tool behavior.

\paragraph{An honest zero.}
On the eight synthetic planted-pattern tasks (T2) the 14B planner never
produced a valid multi-operator mining plan: coverage on that family
was zero, so accuracy was zero while no unsupported claim was emitted.
Abstention is the designed failure mode, but it is a coverage failure,
not a success.

\paragraph{Replication of the development result.}
The frozen CollegeMsg result, 0.408, closely matches the 0.409
development-split result obtained during protocol design (22 tasks,
same configuration), indicating that the development figure was not an
artifact of split selection.

\subsection{Mechanism ablations and verifier validation}
\label{sec:mechanisms}

\paragraph{Output contracts (H2).}
With output-contract checking disabled on the development split,
first-emission ``validity'' \emph{rises} --- the validator simply
checks less --- while execution success and accuracy fall: at 7B,
validity 0.23$\to$0.41 but execution success 0.55$\to$0.23 and accuracy
0.136$\to$0.091; at 14B, validity 0.50$\to$0.55 but accuracy
0.409$\to$0.318. Both runs were repeated at an equal generation budget
with identical results. The check does not create or remove failures;
it moves them from silent runtime deaths to statically repairable
rejections.

\paragraph{Completeness tracking (H3).}
Disabling truncation-taint propagation on the development split lets 3
answers present a fully supported claim over truncated evidence (versus
0 with tainting); most development tasks fit one result page, which is
why the controlled generator below isolates the mechanism more sharply.

\paragraph{Verifier fault injection.}
Table~\ref{tab:mutations} summarizes all mutation classes. The classic
classes perturb mechanically derived, fully grounded answers from
oracle plans (87-answer pool; zero false positives). The
database-semantics classes run on a dedicated non-frozen suite with an
enriched 202-answer pool (zero false positives); a receipted rerun on
the frozen-campaign store reproduces every rate. The wrong-belief-state
class takes its incorrect value from a control execution of the same
plan with \texttt{as\_of\_tt} removed; the truncated-count class
shrinks one page limit so a correct count is computed over a partial
page.

\begin{table}[t]
\centering\small
\caption{Verifier mutation classes. ``Detected'' means the mutated
claim no longer verifies as fully supported. The final column disables
the mechanism under test where one exists.}
\label{tab:mutations}
\begin{tabular}{lrrr}
\toprule
mutation class & cases & detected & w/o mechanism \\
\midrule
count $\pm 1$ & 250 & 250 & --- \\
entity substitution (fabricated) & 250 & 250 & --- \\
real-but-uncited entity added & 100 & 100 & --- \\
ordering operand swap & 100 & 100 & --- \\
unit confusion ($\mu$s as ms) & 100 & 100 & --- \\
wrong belief state & 60 & 60 & --- \\
truncated-page count & 15 & 15 & 0 \\
wrong-step citation & 100 & 36 & --- \\
entity member dropped & 100 & 0 & --- \\
\bottomrule
\end{tabular}
\end{table}

Two entries are deliberate negatives. Wrong-step citations are caught
only when the surrogate step cannot ground the claim (a provenance
pointer naming an uncited step is now rejected outright; the remaining
64 cases were genuinely grounded by the surrogate evidence). Dropped
entity-set members are invisible to trace grounding by construction:
the verifier checks that claimed identifiers appear in cited evidence
(precision), while set completeness is checkable only against the
database --- gold scoring catches it, the trace verifier does not.
The reasoning chain for the truncation row: value-only verification
accepts a count that is arithmetically correct over the returned page;
completeness metadata records that the page is partial; propagating
that metadata through dependent steps prevents the count from being
treated as a complete answer.

\subsection{Model and serving sensitivity}
\label{sec:sensitivity}
These post-campaign studies were run under a different serving
configuration --- a multi-GPU cluster (RTX 5000/6000 Ada, H100; the
University of Memphis iTiger infrastructure~\cite{sharif2025itiger}),
full-precision weights except where noted, vLLM 0.11/cu126, one seed
--- against the same frozen splits and a byte-identically replayed
canonical store. Absolute values should not be compared with the AWQ
primary evaluation; the object is within-configuration trends.

\begin{table}[t]
\centering\small
\caption{Model-scale study. All rows: iTiger cluster, vLLM 0.11/cu126,
temperature 0, one seed, frozen CollegeMsg test split; fp16 except the
72B row (AWQ 4-bit). Emitted unsupported-claim rate is 0.000 in every
row.}
\label{tab:scale}
\begin{tabular}{lccccc}
\toprule
 & \multicolumn{2}{c}{TGMS} & Static-graph & Text-to-Cypher \\
model & acc. & probes & acc. & acc. \\
\midrule
Qwen2.5-7B fp16 & 0.138 & 0.38 & 0.096 & 0.096 \\
Qwen2.5-14B fp16 & 0.340 & 0.77 & 0.032 & 0.191 \\
Qwen2.5-32B fp16 & \textbf{0.628} & \textbf{1.000} & 0.074 & 0.277 \\
Qwen2.5-72B AWQ & 0.511 & 0.31 & 0.096 & 0.138 \\
\bottomrule
\end{tabular}
\end{table}

\paragraph{Scale.}
Under the evaluated prompts, models, and representations, the
operator-backed interface benefits more strongly from increasing
full-precision model size than the two tested baselines (0.138 to 0.628
from 7B to 32B, with probe questions reaching 1.000 at 32B, while
neither baseline exceeds 0.277). This suggests that the constrained
plan representation makes additional planning capability more usable;
it does not establish that serialized-context interfaces categorically
cannot benefit from stronger models. The 14B full-precision row (0.340)
differs from the primary 14B-AWQ result (0.408); the two were served on
different hardware, engines, and precisions and are not directly
comparable.

\paragraph{Quantization.}
The 72B-AWQ configuration underperforms the 32B full-precision
configuration on both accuracy and probes. This result is consistent
with quantization-related degradation in structured planning, but the
present experiment does not separate quantization from model size and
serving effects; a matched pair (32B fp16 vs.\ 32B AWQ, or 72B fp16
vs.\ 72B AWQ) is required to isolate the cause.

\paragraph{Long-context retrieval control.}
Two vector-RAG configurations must be distinguished. The
\emph{original chunking target} ($k{=}20$ chunks of 256 events,
roughly $2\times 10^5$ prompt tokens on this corpus) is not runnable on
any serving window we tested; the primary campaign therefore used
$k{=}1$ of 256 events, and a measured $k{=}2$ request of 36{,}956
tokens was rejected by the serving window. The \emph{long-context
control} ($k{=}20$ chunks of 24 events, ${\sim}$27k tokens, run on an
H100 within the model's native 32k window, 14B fp16) scored 0.021,
consuming 32{,}475 tokens per task, against TGMS's 0.362 at 6{,}521
tokens in the same run. Increasing retrieval breadth from the locally
feasible configuration to this long-context configuration did not
improve accuracy; the main result is therefore not explained solely by
the small $k$ of the original experiment. Chunking granularity,
ranking, serialization, and prompt design were not exhaustively
explored.

\paragraph{Cross-family transfer.}
With the same untuned operator manual, Llama-3.1-8B reaches 0.043
(verified insensitive to generation budget) and Phi-4-mini (3.8B)
0.015. The two smaller untuned models produce few successful plans,
indicating that portability to weaker model families is not automatic;
determining whether a family-independent capability threshold exists
requires matched-size experiments. Across the tested configurations ---
four scales, three families, two quantizations, two serving clusters
--- gated claim support remained stable (no measured unsupported gated
claims) while answer accuracy and coverage varied substantially.

\paragraph{Constrained decoding.}
Grammar-constrained JSON decoding of the plan IR, evaluated as the
escalation for syntactically invalid plans, degraded the 14B
development result on every axis (accuracy 0.409$\to$0.227,
first-emission validity 0.50$\to$0.32, 65\% more wall time): syntactic
validity was not the bottleneck, and post-generation checking with a
repair loop was both safer and cheaper than constraining generation.

\subsection{Operator performance}
\label{sec:opperf}
Operators are benchmarked in isolation on the host CPUs of the
development server (two Xeon Silver 4210, 40 hardware threads, 94\,GB
RAM, DuckDB backend) over synthetic stores of $10^5$, $10^6$, and
$10^7$ edge versions, with one untimed warm-up and seven timed
repetitions; values are operator-only p50 wall times. Two-hop
snapshots run in 13/99/272\,ms across the three scales, global diffs
in 23/163/485\,ms, and metric time series in 144/157/1127\,ms; the
$10^7$ store occupies 2.6\,GB. Over-budget traversals refuse with a
narrowing suggestion rather than degrade. Prompt size stays
${\sim}$6--10k tokens per task at every data scale because structure
never enters the context window.

\section{Design lessons and their experimental support}
\label{sec:findings}
Each lesson follows the same chain: observed failure $\to$ missing
system property $\to$ mechanism added $\to$ controlled test $\to$
consequence.

\paragraph{Lesson 1: models invent \emph{output} fields.}
Our first live run scored 0 of 22 development tasks on execution ---
with every plan passing input validation. Plans referenced
\texttt{s2.count} where the operator emits \texttt{rows\_total}:
argument schemas alone are insufficient for a model-generated dataflow
plan, because nothing constrained what the model read \emph{back}. TGMS
added declared output fields to the operator registry and static
checking of every result reference, with the legal field list in the
repair payload. In the controlled re-run, execution success on the
three probe tasks rose from 0/3 to 3/3 and overall from 0/22 to 12/22;
the effect replicates at 14B (Section~\ref{sec:mechanisms}). Result
schemas deserve the same rigor as argument schemas.

\paragraph{Lesson 2: verification must check evidence completeness.}
A live 14B run called reachability, received the default page of 100
rows (of 343), and computed a flawlessly correct count --- of the page.
Value-only verification accepts that claim; it is false as a statement
about the database. TGMS records completeness metadata on every result
page and propagates truncation taint through dependent steps; claims
resting on tainted evidence are capped below full support. The
controlled generator in Section~\ref{sec:mechanisms} shows the
mechanism is load-bearing: 15/15 truncated-page counts refused full
support with tainting, 0/15 without. The same treatment should extend
to sampling, approximation, timeouts, and stale replicas --- database
answers already know when they are partial.

\paragraph{Lesson 3: serving limits shape baseline design.}
Numeric-dense event text tokenizes at roughly 0.65 tokens per
character, so the original vector-RAG chunking target could not fit any
window we could serve locally, and the primary campaign ran it at
$k{=}1$. When later hardware allowed a runnable $k{=}20$ long-context
control, added breadth did not improve the baseline
(Section~\ref{sec:sensitivity}). Baseline configurations should be
reported together with the serving limits under which they ran, and
retrieval-breadth conclusions should be tested rather than assumed.

\section{Related work}

\paragraph{Temporal and bi-temporal graph databases.}
The distinction between valid time and transaction time is
classical~\cite{snodgrass1999}. Recent systems add temporal support to
property graphs. AeonG~\cite{aeong2024} uses a current and historical
storage split. Gradoop's TPGM supports distributed temporal graph
analysis~\cite{rost2022gradoop}. T-GQL proposes a temporal graph query
language~\cite{debrouvier2021tgql}. These systems are designed mainly for
human-written queries in a general language. TGMS instead exposes a
small operator algebra with explicit types, bounds, costs, and output
fields. These contracts allow an LLM to construct plans and allow the
system to verify their execution. We are not aware of prior work that
combines agent-facing temporal operators, transaction-time reasoning,
and trace-based answer checking in one graph management system.

\paragraph{Graph-augmented retrieval and agent memory.}
GraphRAG summarizes corpus-derived graphs for query-focused
retrieval~\cite{edge2024graphrag}. HippoRAG uses graph structure for
long-term memory consolidation~\cite{gutierrez2024hipporag}.
Zep/Graphiti uses a bi-temporal knowledge graph for agent
memory~\cite{zep2025}, and TOKI formalizes bitemporal write-time
operators for contradiction resolution in relational agent
memory~\cite{wang2026toki}. These systems manage what the agent
remembers or retrieve graph-derived content into the model context. TGMS takes a different approach. The model does not
receive the full graph structure. Operators compute over the graph, and
the model composes the calls and reports the result. TGMS also
quarantines summaries when later corrections overlap their source
windows. TGMS shares the bi-temporal foundation of Zep and TOKI but
targets temporal graph analytics: the checked artifacts are operator
plans and final answer claims rather than memory writes.

\paragraph{Tool use, planning, and program-aided reasoning.}
Toolformer~\cite{schick2023toolformer}, ReAct~\cite{yao2023react}, and
LLMCompiler~\cite{kim2024llmcompiler} show that language models can call
and compose external tools. Program-aided methods delegate arithmetic
and symbolic computation to an interpreter~\cite{gao2023pal,chen2023pot}.
ToolGate attaches Hoare-style pre- and postconditions to individual tool
invocations~\cite{liu2026toolgate}. TGMS applies these ideas to temporal
graphs. Its plan is a constrained DAG over a fixed set of independently
tested operators, checked statically before execution rather than call
by call. The executable surface also enforces identifier grounding and
output-field validity.
Our output-contract result suggests that result schemas deserve the same
care as argument schemas when tools are designed for small models.

\paragraph{Natural-language interfaces to databases.}
Text-to-SQL and related natural-language database interfaces map a
question directly to a query~\cite{yu2018spider}. Our text-to-Cypher
baseline follows this pattern and uses the same models, repair budget,
and answer format as TGMS. TGMS adds a finer-grained audit trail. Each
claim points to named operator results and their digests, rather than
relying only on re-execution of the generated query. Temporal KGQA
benchmarks such as CronQuestions~\cite{saxena2021cron} evaluate
time-aware questions over a fixed knowledge graph. Our correction probes
instead test whether a system can distinguish a past belief state from
the current corrected state.

\paragraph{Faithfulness and claim verification.}
RARR~\cite{gao2023rarr}, FActScore~\cite{min2023factscore}, and
Chain-of-Verification~\cite{dhuliawala2023cove} compare generated text
with retrieved or model-produced evidence; a recent survey frames such
mechanisms as evidence tracing over execution
provenance~\cite{wang2026traces}. TGMS uses narrower evidence with
stronger structure. Claims cite deterministic operator results, so
supported values can be checked exactly. The executor also records
whether evidence was truncated, which prevents a correct value computed
over an incomplete page from being treated as fully supported.

\paragraph{Temporal graph analysis.}
TGMS follows standard temporal-network definitions for time-respecting
paths~\cite{wu2014path,holme2012temporal} and $\delta$-temporal
motifs~\cite{paranjape2017motifs}. The evaluation uses data from the
SNAP and temporal graph benchmark tradition~\cite{panzarasa2009,huang2023tgb}.
TGMS places these algorithms behind bounded operator contracts and
returns an explicit refusal when a request exceeds the cost limit. For
reachability with a maximum-wait constraint, the system uses exact
multi-label search instead of the processing-order-dependent greedy rule
found during specification testing.

\paragraph{Positioning.}
TGMS combines three elements. It provides database-level temporal
semantics with a transaction-time axis. It exposes an agent-facing
computation surface with machine-checkable contracts. It verifies final
claims against deterministic execution evidence. The correction probes
test the first element, the output-contract experiment motivates the
second, and fault injection evaluates the third. The tool server uses
MCP~\cite{mcp2024}, so it can be connected to agent frameworks that
support the protocol.

\section{Limitations}
\label{sec:limits}
\textbf{Workload alignment.} The tasks and the operator set were
co-designed; the study does not yet establish coverage of independently
collected temporal-graph questions.
\textbf{Domain scope.} The real datasets are communication networks;
cross-domain generalization is untested.
\textbf{Capability versus fairness.} Correction probes demonstrate the
value of transaction-time information but compare systems with
different representational capabilities
(Table~\ref{tab:baselineinfo}).
\textbf{Coverage trade-off.} Zero unsupported gated claims is partly
achieved through abstention; coverage and conditional accuracy are
reported alongside it.
\textbf{Verification scope.} Pattern claims are checked but not gated;
approximate operators and probabilistic entity resolution are
unsupported; the guarantees do not extend to adversarial tool behavior
or write-back.
\textbf{Causal uncertainty.} The scale study does not isolate
quantization from size and serving effects, and two small cross-family
models cannot identify a general capability threshold.
\textbf{System maturity.} Concurrency, isolation, incremental index
maintenance, cost-based plan rewriting, and agent write-back remain
open; write-back is schema-ready but disabled pending provenance and
authorization policies.
\textbf{Baseline scope.} The study evaluates particular vector-RAG,
static-graph RAG, and text-to-Cypher implementations, not every
retrieval or query-generation design.

\section{Conclusion}
TGMS explores a database architecture in which an LLM plans and
reports, while the data system owns temporal semantics, deterministic
computation, bounded execution, and evidence checking. The frozen
evaluation shows that this separation is particularly valuable for
questions that require transaction-time belief states and for answers
whose support depends on complete intermediate results. The experiments
also expose two general interface requirements: machine-generated plans
need explicit output contracts, and claim verification must represent
evidence completeness rather than checking values alone. TGMS does not
yet solve general temporal graph question answering: its workload
remains domain-specific, complex multi-operator plans frequently lead
to abstention, and several claim types and approximate operators remain
outside the gating boundary. These limitations motivate cost-based plan
rewriting, broader workload coverage, concurrency control, and
verification with approximate evidence.

\paragraph{Artifacts.}
Code, benchmark generator, task suites, guided demo, and trace viewer:
\url{https://github.com/zxf-work/tgms} (Apache-2.0).

\bibliographystyle{plain}
\bibliography{references}

@inproceedings{paranjape2017motifs,
  author    = {Paranjape, Ashwin and Benson, Austin R. and Leskovec, Jure},
  title     = {Motifs in Temporal Networks},
  booktitle = {Proceedings of the Tenth ACM International Conference on Web
               Search and Data Mining (WSDM)},
  year      = {2017},
  pages     = {601--610}
}

@article{wu2014path,
  author  = {Wu, Huanhuan and Cheng, James and Huang, Silu and Ke, Yiping and
             Lu, Yi and Xu, Yanyan},
  title   = {Path Problems in Temporal Graphs},
  journal = {Proceedings of the VLDB Endowment},
  volume  = {7},
  number  = {9},
  pages   = {721--732},
  year    = {2014}
}

@article{aeong2024,
  author  = {Hou, Jiamin and Zhang, Zhanhao and Wang, Zhouyu and Zhang, Yongjun
             and Lu, Wei and Pan, Anqun and Du, Xiaoyong},
  title   = {AeonG: An Efficient Built-in Temporal Support in Graph Databases},
  journal = {Proceedings of the VLDB Endowment},
  volume  = {17},
  number  = {6},
  year    = {2024},
  pages   = {1515--1527}
}

@misc{zep2025,
  author = {Rasmussen, Preston and Paliychuk, Pavlo and Beauvais, Travis and
            Ryan, Jack and Chalef, Daniel},
  title  = {Zep: A Temporal Knowledge Graph Architecture for Agent Memory},
  year   = {2025},
  note   = {arXiv:2501.13956}
}

@book{snodgrass1999,
  author    = {Snodgrass, Richard T.},
  title     = {Developing Time-Oriented Database Applications in SQL},
  publisher = {Morgan Kaufmann},
  year      = {1999}
}

@article{panzarasa2009,
  author  = {Panzarasa, Pietro and Opsahl, Tore and Carley, Kathleen M.},
  title   = {Patterns and Dynamics of Users' Behavior and Interaction:
             Network Analysis of an Online Community},
  journal = {Journal of the American Society for Information Science and
             Technology},
  volume  = {60},
  number  = {5},
  pages   = {911--932},
  year    = {2009}
}

@inproceedings{kuzu2023,
  author    = {Feng, Xiyang and Jin, Guodong and Chen, Ziyi and Liu, Chang and
               Salihoğlu, Semih},
  title     = {K\`uzu Graph Database Management System},
  booktitle = {Conference on Innovative Data Systems Research (CIDR)},
  year      = {2023}
}

@inproceedings{duckdb2019,
  author    = {Raasveldt, Mark and M{\"u}hleisen, Hannes},
  title     = {DuckDB: An Embeddable Analytical Database},
  booktitle = {Proceedings of the 2019 International Conference on Management
               of Data (SIGMOD)},
  year      = {2019},
  pages     = {1981--1984}
}

@misc{mcp2024,
  author       = {{Anthropic}},
  title        = {Model Context Protocol},
  year         = {2024},
  howpublished = {\url{https://modelcontextprotocol.io}}
}

@article{rost2022gradoop,
  author  = {Rost, Christopher and Gomez, Kevin and Täschner, Matthias and
             Fritzsche, Philip and Schons, Lucas and Christ, Lukas and
             Adameit, Timo and Junghanns, Martin and Rahm, Erhard},
  title   = {Distributed temporal graph analytics with {GRADOOP}},
  journal = {The VLDB Journal},
  volume  = {31},
  pages   = {375--401},
  year    = {2022}
}

@article{debrouvier2021tgql,
  author  = {Debrouvier, Ariel and Parodi, Eliseo and Perazzo, Mat{\'\i}as and
             Soliani, Valeria and Vaisman, Alejandro},
  title   = {A model and query language for temporal graph databases},
  journal = {The VLDB Journal},
  volume  = {30},
  pages   = {825--858},
  year    = {2021}
}

@article{holme2012temporal,
  author  = {Holme, Petter and Saram{\"a}ki, Jari},
  title   = {Temporal networks},
  journal = {Physics Reports},
  volume  = {519},
  number  = {3},
  pages   = {97--125},
  year    = {2012}
}

@inproceedings{huang2023tgb,
  author    = {Huang, Shenyang and Poursafaei, Farimah and Danovitch, Jacob and
               Fey, Matthias and Hu, Weihua and Rossi, Emanuele and
               Leskovec, Jure and Bronstein, Michael and Rabusseau, Guillaume and
               Rabbany, Reihaneh},
  title     = {Temporal Graph Benchmark for Machine Learning on Temporal Graphs},
  booktitle = {Advances in Neural Information Processing Systems (NeurIPS),
               Datasets and Benchmarks Track},
  year      = {2023}
}

@misc{edge2024graphrag,
  author = {Edge, Darren and Trinh, Ha and Cheng, Newman and Bradley, Joshua and
            Chao, Alex and Mody, Apurva and Truitt, Steven and Larson, Jonathan},
  title  = {From Local to Global: A Graph {RAG} Approach to Query-Focused
            Summarization},
  year   = {2024},
  note   = {arXiv:2404.16130}
}

@inproceedings{gutierrez2024hipporag,
  author    = {Guti{\'e}rrez, Bernal Jim{\'e}nez and Shu, Yiheng and Gu, Yu and
               Yasunaga, Michihiro and Su, Yu},
  title     = {{HippoRAG}: Neurobiologically Inspired Long-Term Memory for
               Large Language Models},
  booktitle = {Advances in Neural Information Processing Systems (NeurIPS)},
  year      = {2024}
}

@inproceedings{schick2023toolformer,
  author    = {Schick, Timo and Dwivedi-Yu, Jane and Dess{\`\i}, Roberto and
               Raileanu, Roberta and Lomeli, Maria and Hambro, Eric and
               Zettlemoyer, Luke and Cancedda, Nicola and Scialom, Thomas},
  title     = {Toolformer: Language Models Can Teach Themselves to Use Tools},
  booktitle = {Advances in Neural Information Processing Systems (NeurIPS)},
  year      = {2023}
}

@inproceedings{yao2023react,
  author    = {Yao, Shunyu and Zhao, Jeffrey and Yu, Dian and Du, Nan and
               Shafran, Izhak and Narasimhan, Karthik and Cao, Yuan},
  title     = {{ReAct}: Synergizing Reasoning and Acting in Language Models},
  booktitle = {International Conference on Learning Representations (ICLR)},
  year      = {2023}
}

@inproceedings{kim2024llmcompiler,
  author    = {Kim, Sehoon and Moon, Suhong and Tabrizi, Ryan and Lee, Nicholas and
               Mahoney, Michael W. and Keutzer, Kurt and Gholami, Amir},
  title     = {An {LLM} Compiler for Parallel Function Calling},
  booktitle = {International Conference on Machine Learning (ICML)},
  year      = {2024}
}

@inproceedings{gao2023pal,
  author    = {Gao, Luyu and Madaan, Aman and Zhou, Shuyan and Alon, Uri and
               Liu, Pengfei and Yang, Yiming and Callan, Jamie and
               Neubig, Graham},
  title     = {{PAL}: Program-aided Language Models},
  booktitle = {International Conference on Machine Learning (ICML)},
  year      = {2023}
}

@article{chen2023pot,
  author  = {Chen, Wenhu and Ma, Xueguang and Wang, Xinyi and Cohen, William W.},
  title   = {Program of Thoughts Prompting: Disentangling Computation from
             Reasoning for Numerical Reasoning Tasks},
  journal = {Transactions on Machine Learning Research},
  year    = {2023}
}

@inproceedings{yu2018spider,
  author    = {Yu, Tao and Zhang, Rui and Yang, Kai and Yasunaga, Michihiro and
               Wang, Dongxu and Li, Zifan and Ma, James and Li, Irene and
               Yao, Qingning and Roman, Shanelle and Zhang, Zilin and
               Radev, Dragomir},
  title     = {Spider: A Large-Scale Human-Labeled Dataset for Complex and
               Cross-Domain Semantic Parsing and Text-to-{SQL} Task},
  booktitle = {Proceedings of the 2018 Conference on Empirical Methods in
               Natural Language Processing (EMNLP)},
  year      = {2018}
}

@inproceedings{saxena2021cron,
  author    = {Saxena, Apoorv and Chakrabarti, Soumen and Talukdar, Partha},
  title     = {Question Answering Over Temporal Knowledge Graphs},
  booktitle = {Proceedings of the 59th Annual Meeting of the Association for
               Computational Linguistics (ACL)},
  year      = {2021}
}

@inproceedings{gao2023rarr,
  author    = {Gao, Luyu and Dai, Zhuyun and Pasupat, Panupong and Chen, Anthony and
               Chaganty, Arun Tejasvi and Fan, Yicheng and Zhao, Vincent Y. and
               Lao, Ni and Lee, Hongrae and Juan, Da-Cheng and Guu, Kelvin},
  title     = {{RARR}: Researching and Revising What Language Models Say,
               Using Language Models},
  booktitle = {Proceedings of the 61st Annual Meeting of the Association for
               Computational Linguistics (ACL)},
  year      = {2023}
}

@inproceedings{min2023factscore,
  author    = {Min, Sewon and Krishna, Kalpesh and Lyu, Xinxi and Lewis, Mike and
               Yih, Wen-tau and Koh, Pang Wei and Iyyer, Mohit and
               Zettlemoyer, Luke and Hajishirzi, Hannaneh},
  title     = {{FActScore}: Fine-grained Atomic Evaluation of Factual Precision
               in Long Form Text Generation},
  booktitle = {Proceedings of the 2023 Conference on Empirical Methods in
               Natural Language Processing (EMNLP)},
  year      = {2023}
}

@misc{dhuliawala2023cove,
  author = {Dhuliawala, Shehzaad and Komeili, Mojtaba and Xu, Jing and
            Raileanu, Roberta and Li, Xian and Celikyilmaz, Asli and
            Weston, Jason},
  title  = {Chain-of-Verification Reduces Hallucination in Large Language
            Models},
  year   = {2023},
  note   = {arXiv:2309.11495}
}

@article{wang2026toki,
  author  = {Ziming Wang},
  title   = {{TOKI}: A Bitemporal Operator Algebra for Contradiction
             Resolution in {LLM}-Agent Persistent Memory},
  journal = {arXiv preprint arXiv:2606.06240},
  year    = {2026}
}

@article{liu2026toolgate,
  author  = {Yanming Liu and Xinyue Peng and Jiannan Cao and Xinyi Wang and
             Songhang Deng and Jintao Chen and Jianwei Yin and Xuhong Zhang},
  title   = {{ToolGate}: Contract-Grounded and Verified Tool Execution for
             {LLM}s},
  journal = {arXiv preprint arXiv:2601.04688},
  year    = {2026}
}

@article{wang2026traces,
  author  = {Yiqi Wang and Jiaqi Zhang and Taotao Cai and Zirui Liu and
             Qingqiang Sun and Zequn Sun and Zhangkai Wu and Manqing Dong and
             Mingkai Zheng and Xuefei Yin and Yanming Zhu},
  title   = {From Agent Traces to Trust: A Survey of Evidence Tracing and
             Execution Provenance in {LLM} Agents},
  journal = {arXiv preprint arXiv:2606.04990},
  year    = {2026}
}

@misc{sharif2025itiger,
      title={Cultivating Multidisciplinary Research and Education on GPU
             Infrastructure for Mid-South Institutions at the University
             of Memphis: Practice and Challenge},
      author={Mayira Sharif and Guangzeng Han and Weisi Liu and Xiaolei Huang},
      year={2025},
      eprint={2504.14786},
      archivePrefix={arXiv},
      primaryClass={cs.DC},
      url={https://arxiv.org/abs/2504.14786},
}

\end{document}